\documentstyle[12pt]{article}

\begin{document}

\newcommand{\Lslash}[1]{ \parbox[b]{1em}{$#1$} \hspace{-0.8em}
                         \parbox[b]{0.8em}{ \raisebox{0.2ex}{$/$} }    }
\newcommand{\Sslash}[1]{ \parbox[b]{0.6em}{$#1$} \hspace{-0.55em}
                         \parbox[b]{0.55em}{ \raisebox{-0.2ex}{$/$} }    }
\newcommand{\Mbf}[1]{ \parbox[b]{1em}{\boldmath $#1$} }
\newcommand{\mbf}[1]{ \parbox[b]{0.6em}{\boldmath $#1$} }
\newcommand{\beq}{\begin{equation}}
\newcommand{\eeq}{\end{equation}}
\newcommand{\beqa}{\begin{eqnarray}}
\newcommand{\eeqa}{\end{eqnarray}}
\newcommand{\skipfields}{\!\!\!\!\! & \!\!\!\!\! &}

\newcommand{\gsim}{\buildrel > \over {_\sim}}
\newcommand{\lsim}{\buildrel < \over {_\sim}}
\newcommand{\limt}{{\buildrel {\scriptstyle{t_i\to -\infty}} \over {t_f\to
                    +\infty}}}
\newcommand{\ie}{{\it ie}}
\newcommand{\ieps}{i\varepsilon}
\newcommand{\eg}{{\it eg}}
\newcommand{\cf}{{\it cf}}
\newcommand{\etal}{{\it et al.}}
\newcommand{\gev}{{\rm GeV}}
\newcommand{\jpsi}{J/\psi}
\newcommand{\ket}[1]{\vert{#1}\rangle}
\newcommand{\bra}[1]{\langle{#1}\vert}
\newcommand{\qpair}{$q\bar q$}
\newcommand{\kv}{\vec k}
\newcommand{\pv}{\vec p}
\newcommand{\fhalf}{\frac{1}{2}}
\newcommand{\half}{{1/2}}
\newcommand{\order}[1]{${\cal O}(#1)$}
\newcommand{\morder}[1]{{\cal O}(#1)}
\newcommand{\eq}[1]{Eq.\ (\ref{#1})}

\newcommand{\ptr}{p_T}
\newcommand{\as}{\alpha_s}

\newcommand{\PL}[3]{Phys.\ Lett.\ {#1} ({#2}) {#3}}
\newcommand{\NP}[3]{Nucl.\ Phys.\ {#1} ({#2}) {#3}}
\newcommand{\PR}[3]{Phys.\ Rev.\ {#1} ({#2}) {#3}}
\newcommand{\PRL}[3]{Phys.\ Rev.\ Lett.\ {#1} ({#2}) {#3}}
\newcommand{\ZP}[3]{Z. Phys.\ {#1} ({#2}) {#3}}
\newcommand{\PRe}[3]{Phys.\ Rep.\ {#1} ({#2}) {#3}}

%
\newcommand{\LevelOne}[1]{ \newpage \subsection*{#1} }
\newcommand{\LevelTwo}[1]{ \subsubsection*{#1} }
\newcommand{\LevelThree}[1]{ \paragraph*{#1} }
\newcommand{\AppendixLevel}[2]{
   \newpage
   \subsection*{Appendix #1 -- #2}
   \markright{Appendix #1}
   \addcontentsline{toc}{subsection}{\protect\numberline{#1}{#2}} }

\begin{titlepage}
\begin{flushright}
        NORDITA-96/63 P
\end{flushright}
\begin{flushright}
        hep-ph/9610270 \\ \today
\end{flushright}
\vskip .8cm
\begin{center}
{\Large A Perturbative Gluon Condensate?\\}
\vskip .8cm
 {\bf Paul Hoyer\footnote{
Email: hoyer@nordita.dk. Work supported in part by the EU/TMR contract ERB
FMRX-CT96-0008.}}
\vskip .5cm
NORDITA\\
Blegdamsvej 17, DK-2100 Copenhagen \O
\vskip 1.8cm
\end{center}

\begin{abstract} \noindent
I propose that the properties of QCD perturbation theory should be
investigated when the boundary state (`perturbative vacuum') at $t=
\pm\infty$ includes gluons. Any boundary state that has an overlap
with the true QCD ground state generates a perturbative series that
(when summed to all orders) is formally exact.
Through an analogy with the boundary condition corresponding to a fermion
condensate, I propose an explicit form for a `perturbative gluon
condensate' that suppresses low momentum gluon production, thus generating
an effective mass gap. Standard perturbative calculations are modified
only through a change in the $\ieps$ prescription of low momentum $(|\pv|
\lsim \Lambda_{QCD})$ gluon propagators. Gauge invariance is expected to be
preserved since this modification is equivalent to adding on-shell
external particles. Renormalizability is unaffected since only low-momentum
propagators are modified. Due to the asymptotic low momentum
gluons boost invariance is not explicit. Lorentz invariance should be
restored in the sum to all orders in analogy to standard bound state
calculations.

\end{abstract}

\end{titlepage}

\setlength{\baselineskip}{7mm}

\subsection*{1. Introduction}

Hadron wave functions appear phenomenologically to be frame
dependent. In the infinite-momentum (or light-cone) frame the proton
is observed to have, in addition to its $uud$ valence quarks, important
gluon and sea quark components \cite{pdg}. In particular, gluons carry about
half of the proton momentum. This measurement of the proton is rather
rigorously justified by the QCD factorization theorem \cite{fact}. The
success of perturbative QCD predictions for a large number of hard scattering
processes has established QCD as the correct theory for the strong
interactions.

The non-relativistic quark model (NRQM) \cite{pdg,nrqm,lipkin} provides a
less rigorous but phenomenologically very successful and simple rest frame
picture of hadrons as non-relativistic bound states of `constituent' quarks.
The masses of the $u,d$ constituent quarks are $\morder{300\ \rm{MeV}}$,
which is considerably larger than the `current' quark masses which are
relevant for short-distance processes. The `missing' gluon and sea quark
degrees of freedom appear to be frozen in the structure of the constituent
quarks. 

The simple regularities of the hadron spectrum, coupled with the
success of QCD as applied to hard processes, obviously invites efforts
to find a QCD justification of the NRQM (see, \eg,
\cite{nrqm,cgip,gri,ris,diak}). The challenge is to find a formulation
which in a first approximation retains the simplicity of the quark model,
yet allows corrections specified by QCD to be evaluated to arbitrary
order.

The similarities of the hadron spectrum with QED bound states, together with
the success of perturbative calculations in that theory, suggests the use of
a perturbation expansion also in QCD. Such an expansion is determined by the
lagrangian and by the boundary conditions at asymptotic times. Central
properties of QCD like the finite range of the color force and
confinement are commonly associated with a non-trivial ground state of the
theory, the `gluon condensate' \cite{con,dos}. In this paper I shall
propose a specific boundary condition on QCD perturbation theory which
is motivated by the gluon condensate and by a suppression of soft
gluon production. The purpose is not to model the condensate in detail. Just
as in QED it may suffice to achieve so much overlap with the true
ground state that low orders of the perturbative series already
incorporates the main physical features of the theory. Corrections are then
given systematically by the higher orders of the expansion.

This work points to a perturbative expansion of QCD which appears not
substantially more difficult to evaluate than the standard one, but which
has a number of novel features. The usefulness of this approach can only be
judged after a further study of the properties of that expansion.

\subsection*{2. A fermion condensate}

The asymptotic states that we impose on perturbative expansions at initial
and final times ($t=\pm\infty$) should have an overlap with the true ground
state of the theory. This guarantees that the full perturbative
expansion formally gives exact results. In euclidean formulations this fact
is particularly clear since the time development of energy eigenstates is
given by $\exp(-E\tau)$, implying a dominance of the true ground
state (of lowest energy $E$) in the limit $\tau \to \infty$. In
minkowski space the same result is obtained using an $\ieps$
prescription\footnote{For a discussion of boundary states in field theory
see, \eg, Ref. \cite{keld}}.

This freedom in the choice of boundary states allows for a whole set of
formally equivalent perturbative expansions. Since all expansions are
expected to diverge, their equivalence is of more formal than practical
significance. From a practical point of view what matters is that the lowest
orders already incorporate the main physical characteristics of the theory.

We have little understanding of the structure of the QCD gluon condensate in
terms of Fock state wave functions. It nevertheless seems plausible that the
vacuum wave function components involving gluons and quarks of 3-momenta
smaller than the characteristic QCD scale $\Lambda_{QCD}$ are strongly
modified. In this respect, the gluon condensate may resemble a fermi
condensate with fermi momentum of \order{\Lambda_{QCD}}. In a fermi
condensate the exclusion principle prevents pair production below the fermi
momentum. It seems desirable to have a similar property for gluons, to
suppress soft gluon production which can give rise to long-range color
correlations.

In this section I recall how perturbation theory is modified in the
presence of a fermion condensate. Only the $\ieps$ prescription is affected
-- which is enough to have significant consequences. I shall then in the
next section use this as a guide for constructing a `perturbative gluon
condensate', namely one that results in an analogous modification of the
$\ieps$ prescription for the gluon propagator. Having shown that there
exists a boundary condition which implies such an $\ieps$ modification for
gluons it can for many practical purposes be forgotten, and the usual
feynman diagrams be evaluated with modified (low momentum) propagators.

The standard free fermion propagator
\beq
iS_F(x-y)= \langle 0|T[\psi(x)\bar\psi(y)]|0\rangle   \label{sfxy}
\eeq
is in momentum space
\beq
S_F(p)= \frac{\Sslash{p}+m}{p^2-m^2+\ieps} =
\frac{\Sslash{p}+m}{(p^0-E_p+\ieps)(p^0+E_p-\ieps)}~~.  \label{sfp}
\eeq
If we add an antifermion to the initial and final states,
\beqa
\langle 0|d_{\lambda'}(\kv')T[\psi(x)\bar\psi(y)]d_{\lambda}^{\dag}
(\kv)|0\rangle &=&
iS_F(x-y)2E_k (2\pi)^3 \delta^3(\kv-\kv')\delta_{\lambda\lambda'} 
 \nonumber \\
&+& v(\lambda',\kv') \bar v(\lambda,\kv) e^{ik'\cdot x - ik\cdot y}~~, 
\label{onef}
\eeqa
the feynman propagator is multiplied by the annihilation amplitude for the
inserted antifermions, and there is a new term corresponding to a mixing
of the antifermion propagating from $x$ to $y$ with the
antifermion in the in- and out-states.

For a condensate we would fill both helicity states at a given
momentum $\kv$. The free propagator
\beq
iS(x-y)\equiv \langle 0|d_\half(\kv)d_{-\half}(\kv)T[\psi(x)\bar
\psi(y)]d_{-\half}^{\dag}(\kv) d_\half^{\dag} (\kv)|0\rangle  \label{twofxy}
\eeq
is then in momentum space
\beq
S(p)= \left[(2\pi)^3 2E_k \delta^3(\vec 0)\right]^2 \left\{
\begin{array}{ll}S_F(p)&\ \ \ \ (\pv \neq -\kv) \\
                 S_E(p)&\ \ \ \ (\pv = -\kv) \\
\end{array} \right.  \label{twofp}
\eeq
where
\beq
S_E(p)= \frac{\Sslash{p}+m}{(p^0-E_k+\ieps)(p^0+E_k+\ieps)}  \label{sep}
\eeq
differs from the feynman propagator only in the $\ieps$ prescription at
$p^0=-E_k$. Since the antifermions inserted in the definition
(\ref{twofxy}) are on-shell, it is clear that a mixing between them and
the propagating fermion only can occur at the antifermion pole of $S(p)$.
Adding antifermions for all momenta $|\kv| \leq \Lambda$, the
corresponding propagator will equal $S_E(p)$ for all $|\pv| \leq \Lambda$.

The addition of (anti)fermions at $t=\pm\infty$ influences the fermion
propagators in feynman diagrams at any order of perturbation theory
exactly as it does the lowest order propagator above. This can be easily
seen using the generating functional of green functions (in a theory like
QED or QCD),
\beq
Z[\zeta,\bar\zeta;J] = \exp\left[ iS_{int}\left(\frac{\delta}{\delta\zeta},
\frac{\delta}{\delta\bar\zeta};\frac{\delta}{\delta J}\right)\right]
Z_B[J]Z_F[\zeta,\bar\zeta]  \label{genfun}
\eeq
where $S_{int}$ is the interaction part of the action and $Z_B,~Z_F$ are
free functionals of the boson $(J)$ and fermion $(\zeta,~\bar\zeta)$
sources, respectively. The free fermion functional is
\beq
Z_F[\zeta,\bar\zeta]=\exp\left[i\int\frac{d^4p}{(2\pi)^4} \bar\zeta(-p)
S_F(p) \zeta(p) \right]  \label{fgenfun}
\eeq
with the feynman propagator given by \eq{sfp}.

It is instructive first to rederive the result (\ref{twofp}) for the free
propagator with the boundary states (\ref{twofxy}) using the generating
functional. The propagator $S_F(p)$ is diagonalized by the sources
$z,~\bar z$ of definite helicity $(\lambda)$ and energy signature $(\pm)$,
\beqa
\zeta(p) &=& \frac{\gamma^0}{\sqrt{2E_p}}\sum_\lambda 
  \left[ u(\lambda,\pv)\,z_+^\lambda(p) + v(\lambda,-\pv)\,z_-^\lambda(p)
  \right] \nonumber \\
\bar\zeta(-p) &=& \frac{1}{\sqrt{2E_p}}\sum_\lambda 
  \left[ \bar z_+^\lambda(-p)\,u^{\dag}(\lambda,\pv) +
         \bar z_-^\lambda(-p)\,v^{\dag}(\lambda,-\pv)
  \right]  \label{zdef}
\eeqa
In the new basis we have
\beq
Z_F[z,\bar z]= \exp\left\{i\int\frac{d^4p}{(2\pi)^4} \sum_\lambda \left[
\frac{\bar z_+^\lambda(-p)z_+^\lambda(p)}{p^0-E_p+\ieps} +
\frac{\bar z_-^\lambda(-p)z_-^\lambda(p)}{p^0+E_p-\ieps} \right]\right\}
\label{zgf}
\eeq
The generating functional $Z_E$ of the modified propagator (\ref{sep})
(for some given $\kv$) differs by the sign of $\ieps$ in the second term
of \eq{zgf}. Hence (I suppress the 3-momentum $\pv$ and factors $(2\pi)^3
2E \delta^3(\vec 0)$ in the following),
\beqa
Z_E[z,\bar z] &=& \exp\left[i\int\frac{dp^0}{2\pi}\, \bar\zeta(-p)
S_E(p) \zeta(p) \right] \nonumber \\
&=& \prod_\lambda\left[ 1+\bar z_-^\lambda(E)z_-^\lambda(-E)
\right] Z_F[z,\bar z]~~,  \label{zegf}
\eeqa
where I used $(p^0+E+\ieps)^{-1}-(p^0+E-\ieps)^{-1}= -2\pi i\delta(p^0+E)$,
and $\exp(\bar z z)= 1+\bar z z$ for grassmann sources $\bar z,z$.

As a function of time,
\beqa
z(p^0)&=&\int dt'\, z(t') e^{it'p^0} \nonumber \\
\bar z(-p^0)&=&\int dt''\, \bar z(t'') e^{-it''p^0}  \label{tspace}
\eeqa
the free generating functionals are
\beqa
Z_F&=&\exp\left\{dt'dt'' \sum_\lambda
\left[\bar z_+^\lambda (t'')\theta(t''-t') e^{-iE(t''-t')} z_+^\lambda(t')
 \right. \right. \nonumber 
\\ && \ \  \left. \left.
-\bar z_-^\lambda (t'')\theta(t'-t'') e^{-iE(t'-t'')} z_-^\lambda(t')
\right] \right\}  \label{tgff} \\
Z_E&=&\prod_\lambda \left[ 1+
\int dt'dt'' \bar z_-^\lambda (t'') e^{-iE(t'-t'')} z_-^\lambda(t')\right]
Z_F  \label{tgfe}
\eeqa
This expression for $Z_E$ can now be compared with the one obtained by
explicitly differentiating $Z_F$ wrt. its sources at $t=\pm\infty$,
corresponding to the boundary states of \eq{twofxy}. One readily finds
\beq
Z_E= \lim_\limt \prod_\lambda \left[ e^{iE(t_f-t_i)}
\frac{\delta^2}{\delta\bar z_-^\lambda(t_i) \delta z_-^\lambda(t_f)}
\right] Z_F  \label{frel}
\eeq

This result extends immediately to the full interactive functional
(\ref{genfun}), since the derivatives in \eq{frel} commute through the
derivatives in $\exp(iS_{int})$. This means that using the modified fermion
propagator $S_E$ of \eq{sep} everywhere in a perturbative calculation of an
arbitrary green function (for some given 3-momentum $\kv$ of the
propagators) is exactly equivalent to calculating the same green function
using ordinary feynman propagators but with additional incoming and outgoing
antifermions as in \eq{twofxy}.

The change of $\ieps$ prescription suppresses fermion pair production at
the corresponding value(s) of $\kv$, as required by the exclusion
principle. This can be seen directly for, \eg, a fermion loop correction
to a gauge boson propagator. The loop gives no contribution at those values
of the fermion momenta $\kv$ at which external fermions have been
introduced, since the poles in the loop momentum $p^0$ then are all below the
real axis and the $p^0$ integral may be closed in the upper half plane.

Since I have shown that the propagator modification is equivalent to
adding particles at $t=\pm\infty$ gauge invariance is likely to be
preserved. Formally, the ward identities involve inverse propagators, for
which the sign of $\ieps$ is irrelevant.

\subsection*{3. A boson `condensate'}

Boundary conditions like that of \eq{twofxy} with (anti)fermions added to
the in- and out-states are relevant in situations involving fermion
condensates, but not for typical applications of QCD. The QCD vacuum has
zero baryon number, and thus no overlap with states having extra
(anti)quarks. The propagator modification nevertheless seems
phenomenologically interesting for gluons, since it suggests a `freezing'
of the low momentum gluon d.o.f.'s. Effective gluon and constituent quark
masses can be generated through loop corrections due to the propagator
modification, presumably without loss of gauge invariance (but with loss of
lorentz invariance order by order, see section 4). 

I shall show that there is a boundary condition
which implies an analogous modification of the
$\ieps$ prescription for boson propagators as the one for fermions
discusses above. Not surprisingly, this `perturbative boson condensate'
involves an indefinite number of external bosons. For simplicity, I shall
consider scalar bosons only. The generalization to real (transverse) gluons
should be straightforward.

The free boson functional appearing in \eq{genfun} is (for scalars)
\beq
Z_B[J]= \exp\left[\frac{i}{2}\sum_{\pm\pv}\int\frac{dp^0}{2\pi}
J(-p^0,-\pv) D_F(p) J(p^0,\pv) \right]  \label{bgf}
\eeq
Since we shall be dealing with the free functional (the generalization to
the interacting one will again be straightforward), it is sufficient to
consider a single 3-momentum $\pv$, and keep only the bose symmetrization
over $\pm\pv$ as indicated in \eq{bgf}. The feynman propagator is
\beq
D_F(p)= \frac{1}{p^2-m^2+\ieps} = \frac{1}{2E}\left(
\frac{1}{p^0-E+\ieps}-\frac{1}{p^0+E-\ieps} \right)  \label{bfp}
\eeq
where $E=\sqrt{\pv^2+m^2}$. A modification of the $\ieps$ prescription at
the $p^0=-E$ pole gives
\beq
D_E(p) \equiv \frac{1}{(p^0-E+\ieps)(p^0+E+\ieps)} =
D_F(p)+ \frac{2\pi i}{2E}\delta(p^0+E)  \label{bep}
\eeq
Note that the same generating functional is obtained if the $\ieps$
prescription is changed instead at the $p^0=+E$ pole. Thus
\beq
\tilde D_E(p) \equiv \frac{1}{(p^0-E-\ieps)(p^0+E-\ieps)} = D_E(-p)
\label{detil}
\eeq
so that
\beq
\sum_{\pm\pv} \int \frac{dp^0}{2\pi} J(-p)\tilde D_E(p) J(p) =
\sum_{\pm\pv} \int \frac{dp^0}{2\pi} J(-p) D_E(p) J(p) \label{gfequiv}
\eeq

In $(t,\pv)$-space,
\beq
J(p^0,\pv)= \int dt\, J(t,\pv) e^{itp^0}  \label{ftj}
\eeq
we have
\beqa
Z_B[J]&=&\exp\left\{\sum_{\pm\pv} \frac{1}{4E}\int dt'dt'' J(t'',-\pv)
\right. \nonumber \\
&\times& \left. \left[ \theta(t''-t')e^{-iE(t''-t')} +
\theta(t'-t'')e^{iE(t''-t')} \right] J(t',\pv) \right\}  \label{ftbgf}
\eeqa
The generating functional for the modified scalar propagator (\ref{bep})
is then
\beqa
Z_E[J] &\equiv& \exp\left[\frac{i}{2}\sum_{\pm\pv} \int\frac{dp^0}{2\pi}
J(-p^0,-\pv) D_E(p) J(p^0,\pv) \right] \label{bedef} \\
 &=& \exp\left[ -\sum_{\pm\pv}\frac{1}{4E} J(E,-\pv)J(-E,\pv) \right]Z_B[J]
\nonumber \\
 &=& \exp\left[ -\sum_{\pm\pv}\frac{1}{4E}\int dt'dt'' J(t'',-\pv)
e^{iE(t''-t')}J(t',\pv) \right] Z_B[J]  \label{ftbegf}
\eeqa

\eq{ftbegf} may be compared with \eq{tgfe} in the fermion case. Due to the
grassmann algebra, the exponential factor multiplying $Z_F$ contains only
a single power of the fermion sources $\bar z,z$. In the boson case the
factor multiplying $Z_B$ in \eq{ftbegf} contains arbitrary powers of the
sources $J$. It can be reproduced only by differentiating $Z_B[J]$ an
arbitrary number of times, corresponding to an indefinite number of
incoming and outgoing bosons.

A single boson of momentum $\pv$ in the in-state is obtained as
\beq
\lim_{t_i\to-\infty} \frac{\delta Z_B[J]}{\delta J(t_i,-\pv)} =
\left[\frac{1}{2E} \int dt\, e^{-iE(t-t_i)}J(t,\pv)\right] Z_B[J]
\label{inb}
\eeq
where we used $\lim_{t_i\to-\infty}\theta(t_i-t)=0$. Similarly an outgoing
boson corresponds to
\beq
\lim_{t_f\to+\infty} \frac{\delta Z_B[J]}{\delta J(t_f,\pv)} =
\left[\frac{1}{2E} \int dt\, e^{-iE(t_f-t)}J(t,-\pv)\right] Z_B[J]
\label{outb}
\eeq
Having a boson both incoming and outgoing is then given by
\beqa
&&\lim_\limt  e^{iE(t_f-t_i)} 2E \frac{\delta^2 Z_B[J]}
{\delta J(t_f,\pv)\delta J(t_i,-\pv)} = \hfill \nonumber \\
 && \ \ \ \left[ 1+\frac{1}{2E} \int dt'dt''\, J(t'',-\pv) e^{iE(t''-t')}
J(t',\pv)\right] Z_B[J]  \label{inoutb}
\eeqa
Further differentiation wrt. $J(t_i,-\pv)$ and $J(t_f,\pv)$ now operates
also on the first factor in \eq{inoutb}. However, this gives back the
factors in Eqs. (\ref{inb}) and (\ref{outb}), respectively. Hence applying
the double derivative of \eq{inoutb} any number of times on $Z_B$
generates a polynomial factor in $xy$, where
\beqa
x &\equiv& \frac{1}{\sqrt{2E}}\int dt''\,J(t'',-\pv) e^{iEt''} \nonumber\\
y &\equiv& \frac{1}{\sqrt{2E}}\int dt'\,e^{-iEt'}J(t',\pv)~~. 
\label{xydef}
\eeqa
According to \eq{ftbegf},
\beq
Z_E[J]= \exp\left(-\fhalf \sum_{\pm\pv} xy \right) Z_B[J]~~. \label{zerel}
\eeq
We need to consider only how to generate the $+\pv$ term in
\eq{zerel} through repeated differentiation of $Z_B$ as in \eq{inoutb}.
The  $-\pv$ term will then be obtained similarly through repeated
$\delta^2/\delta J(t_f,-\pv)\delta J(t_i,\pv)$ differentiation.

The function
\beq
f(xy) \equiv \exp \left(\lambda\frac{\partial^2}{\partial x\partial y}
\right) \exp(xy)  \label{fxydef}
\eeq
provides an adequate model for the present problem. As seen from
\eq{ftbgf}, $Z_B$ is not of the form $\exp(xy)$ due to the
$\theta$-functions, but as in Eqs. (\ref{inb}~--~\ref{inoutb}) $Z_B$ acts
precisely like 
$\exp(xy)$ when differentiated in the limits $t_i\to -\infty,\ 
t_f\to +\infty$. Hence the polynomial in $xy$ generated by the derivatives
in \eq{fxydef} will be the same as that generated from $Z_B$.

It is straightforward to evaluate $f(xy)$ in \eq{fxydef} by using the
identity
\beq
\exp\left(\fhalf x^2 \right) = \frac{1}{\sqrt{2\pi}}
\int_{-\infty}^\infty du \exp\left(-\fhalf u^2+ux \right)  \label{intrep}
\eeq
to express
\beq
\exp(xy)= \exp\left[\fhalf(x+y)^2\right] \exp\left(-\fhalf x^2 \right)
\exp\left(-\fhalf y^2 \right) \label{erep}
\eeq
as a three-fold integral. The integral resulting from applying the
derivatives in \eq{fxydef} is gaussian and gives
\beqa
\exp \left(\lambda\frac{\partial^2}{\partial x\partial y}\right) \exp(xy)
&=& \frac{1}{1-\lambda} \exp\left(\frac{xy}{1-\lambda}\right) \nonumber \\
&=& \frac{1}{1-\lambda} \exp\left(\frac{\lambda}{1-\lambda}xy\right)
\exp(xy) \label{fres}
\eeqa
Requiring that the first exponent in \eq{fres} be $-xy/2$ according to
\eq{zerel} gives the result $\lambda=-1$.

We have thus shown that the generating functional $Z_E[J]$ (\ref{bedef})
of the scalar propagator $D_E(p)$ (\ref{bep}), which differs from the
feynman propagator $D_F(p)$ (\ref{bfp}) by the sign of $\ieps$ at the
$p^0=-E$ pole, is equivalent to the standard generating functional $Z_B[J]$
(\ref{bgf}) of feynman propagators differentiated wrt. sources at
$t=\pm\infty$,
\beq
Z_E[J]=4\exp\left[-\sum_{\pm\pv}\lim_\limt e^{iE(t_f-t_i)} 2E
\frac{\delta^2} {\delta J(t_f,\pv)\delta J(t_i,-\pv)} \right] Z_B[J] 
\label{zeres}
\eeq
As noted in \eq{gfequiv}, the same result obtains if the $\ieps$
prescription is modified at the $p^0=+E$ pole instead.

The source derivatives in (\ref{zeres}) commute through the interaction
term in the definition (\ref{genfun}) of the full generating functional of
green functions in the interacting theory. Hence the above result
establishes that a perturbative calculation (to arbitrary order) which
uses the propagator $D_E(p)$ (\ref{bep}), with its non-standard $\ieps$
prescription, is equivalent to a standard perturbative calculation using
feynman propagators in the presence of a `perturbative condensate' of
incoming and outgoing particles as specified by the source derivatives in
\eq{zeres}.

\subsection*{4. Discussion}

I have argued that it may be useful to consider perturbative expansions of
QCD using non-trivial boundary conditions at $t=\pm\infty$, given that the
ground state of the theory is a gluon condensate. All expansions in which the
boundary states overlap the true ground state are formally equivalent and
{\em a priori} equally good.

I investigated a particular case which is the bosonic equivalent of a
fermion condensate, and in which the $\ieps$ prescription of low momentum
boson propagators is modified. Such a propagator modification corresponds
to a superposition of standard perturbative calculations where 0, 1, 2, etc.
bosons are added both to the initial and final state, as
expressed by \eq{zeres}. I have not shown that these boundary states
have an overlap with the true QCD vacuum (but then, neither do we know
that the standard perturbative vacuum has such an overlap).

The relevance of this expansion depends on its theoretical and
phenomenological viability, which remains to be demonstrated. Gauge
invariance is among the important properties that should be explicitly
verified.

Since only low-momentum $(|\pv| \lsim \Lambda_{QCD})$ propagators are
modified, the successful results of `hard' QCD processes remain
unaltered. In particular, the renormalization procedure will not be
affected in any way by the modifications suggested here.

The most striking difference compared to standard perturbation theory is
the lack of boost invariance order by order. Contrary to what
might first appear, this need not signal a breakdown of lorentz symmetry
for the full series. The true asymptotic degrees of freedom are the hadron
bound states, which do not occur at any finite order of perturbation
theory. Physical symmetry requirements should be imposed only on
resummations of the series.

The subtleties of lorentz invariance in bound state calculations is
known from QED. As an example \cite{lep}, consider the lippman-schwinger
equation
\beq
G_T(E) = K(E)+K(E)S(E)G_T(E)  \label{dys}
\eeq
for the (truncated) green function $G_T$ of a $2\to 2$ process with c.m.
energy $E$. Iterating this equation generates an expansion of $G_T$
in powers of the propagator $S$ and the kernel $K$. While the standard
perturbative expansion in $\alpha$ is unique (up to renormalization
conventions) for the green function $G_T$, this is not so for
$S$ and $K$ separately. Rather, we can choose the form of the propagator $S$
freely, be it of relativistic (dirac) or non-relativistic (schr\"odinger)
form. \eq{dys} then determines the corresponding perturbative expansion of
the kernel $K$. At a pole of the (full) green function of the form
\beq
G(E) = \frac{\psi_n \bar\psi_n}{E-E_n} + \mbox{regular terms}
\label{pol}
\eeq
the lippman-schwinger equation implies a bound state equation of the form
\beq
S^{-1}(E_n) = K(E_n) \psi_n~~.  \label{dbse}
\eeq
For a non-relativistic propagator $S$ this will have the form of a
schr\"odinger equation, but it will give exact results provided the full
perturbative series for the interaction kernel $K$ is used.

It should furthermore be realized that the transformation properties of
equal-time bound state wave functions under lorentz boosts is quite
non-trivial. The requirement that the constituents should be evaluated at
equal time in all frames is inconsistent with explicit space-time
covariance, even for non-relativistic QED bound states.

A simple example serves to illustrate the novel aspects of
the frame dependence of equal-time wave functions. There is a bound state
equation in QED$_2$ for which it is possible to relate explicitly the
solutions in different lorentz frames, and thus verify that they have the
correct transformation properties \cite{boo}. The wave
function of a two fermion bound state is written 
\beq 
\psi(t,x_1,x_2) = \exp(-iEt)\, \exp\left(ik\frac{x_1+x_2}{2}\right)\,
\chi(x_1-x_2)~~, 
\label{wf} 
\eeq 
where $x_1,x_2$ are the positions of the constituents and
$t$ their common time. Both the bound state energy $E$ and the $2\times 2$
dirac wave function $\chi$ depend on the bound state c.m. momentum
parameter $k$. The bound state equation for $\chi$ is 
\beq
-i\partial_x \left[\alpha,\chi(x)\right] +\fhalf k 
\left\{\alpha,\chi(x)\right\} +m_1\gamma^0 \chi(x) -m_2\chi(x) \gamma^0
= \left(E-V(x)\right) \chi(x)  \label{bse} 
\eeq
where $m_1,m_2$ are the constituent masses and $V(x)=\fhalf e^2 |x|$ is the
instantaneous Coulomb potential. In
1+1 dimensions we may represent the dirac matrices using pauli
matrices, $\gamma^0=\sigma_3$ and $\alpha = \gamma^0\gamma^1 =
\sigma_1$. Despite the fact that \eq{bse} has no explicit lorentz
covariance (space and time coordinates are treated differently in Eqs.
(\ref{wf},\ref{bse})) the bound state energies for different c.m. momenta
$k$ are correctly related: $E=\sqrt{k^2+M^2}$, with $M$ independent of $k$.
In the limit of non-relativistic internal motion\footnote{This is in fact the
only case where the solutions are normalizable and thus meaningful, due to
the Klein paradox. The bound state momentum $k$ can be arbitrarily large,
however.}
$(e/m_{1,2} \ll 1)$ the wave function $\chi(x)$  lorentz contracts in the
standard way as a function of $k$. Related examples may be found in Ref.
\cite{lor}.

The ground state wave function of QCD$_4$ is invariant under boosts.
This is obviously not the case for our asymptotic states which
according to \eq{zeres} contain bosons of definite 3-momenta ($\lsim
\Lambda_{QCD}$). In the present formulation, the same boundary states must
be used in all frames, since they model the same (invariant) ground state.
Hence the perturbative expansion of a given QCD process will depend on the
lorentz frame. If the method works, measurable quantities such as hadronic
cross sections will be lorentz invariant. This does
not mean that the hadron wave functions themselves will be invariant --
the starting point of this paper was in fact that they appear
phenomenologically to be strongly frame dependent.

\bigskip\noindent
{\bf Acknowledgements.} I would like to thank V. M. Braun and S. J. Brodsky
for useful discussions.


\end{document}